# Nanofilm Materials for Devices of Magnetic Field Measurement in Radiation Environment

I. Bolshakova, P. Horelkin, Ya. Kost, A. Moroz,
Y. Mykhashchuk, M. Radishevskiy, F. Shurigin,
O. Vasyliev
Magnetic Sensor Laboratory
Lviv Polytechnic National University
Lviv, Ukraine
inessa@mail.lviv.ua

B. Pavlyk
Department of Sensory and Semiconductor Electronics
Ivan Franko National University of Lviv
Lviv, Ukraine
pavlyk@electronics.lnu.edu.ua

T. Kuech
College of Engineering
University of Wisconsin-Madison
Madison, Wisconsin, USA
tfkuech@wisc.edu

Z. Wang, M. Otto, D. Neumaier
Advanced Microelectronic Center Aachen (AMICA)
AMO GmbH
Aachen, Germany
wang@amo.de

*Abstract*—The prospects of using nanofilms of indium-containing III-V semiconductors, gold and single-layer graphene in magnetic field sensors, intended for application in radiation environment were evaluated on the results of testing in neutron fluxes. Semiconductor sensors are capable of withstanding radiation levels typical for the ITER-type fusion reactors, while gold sensors are stable even under environment expected in the first fusion power plant DEMO. Graphene is promising for creating sensors that combine high magnetic field sensitivity and high irradiation resistance.

## I. INTRODUCTION

Hall sensors are the most common tools for magnetic measurement in many areas of science and technology, due to their ability to operate in a wide range of the magnetic field's induction and frequency, as well as the design simplicity, small sizes and low cost [1]. Today, most of these sensors are made on the basis of silicon, which provides a sufficiently high sensitivity to the magnetic field, but quickly degrades under radiation conditions [2]. This fact limits the use of industrial Hall sensors in such responsible systems as nuclear power plants, spacecrafts, charged particle accelerators, etc., where magnetic measurements can be a powerful instrument for the efficient control of technological processes and the safety enhancement.

In particular, Hall sensors allow to improve significantly an accuracy of the plasma's magnetic diagnostics in steady-state fusion reactors [3], among which are the experimental tokamak ITER that is under construction now in France, as well as the first fusion power plant DEMO that is designed by the European consortium EUROfusion and will start to bring the energy into grid already by mid-century. In such facilities, the magnetic diagnostics of plasma is a critical tool for its parameters control and its holding at a safe distance from the reactor's construction elements. At the same time, extremely high neutron fluences are expected in the locations of magnetic sensors over their operation time – up to $1.3 \times 10^{22}$ n·m$^{-2}$ (ITER) and $(1 - 100) \times 10^{24}$ n·m$^{-2}$ for DEMO [4, 5]. This stimulates the search for new materials for Hall sensors, which are capable of maintaining the electrophysical characteristics' stability for a long time under the influence of high neutron fluxes and high temperatures.

The goal of work is to evaluate the prospects of application of modern nanofilm materials – the indium-containing III-V semiconductor nanoheterostructures, gold nanofilms and single-layer graphene – in the irradiation-resistant Hall sensors intended for operation in the environment with high neutron fluences.

## II. EXPERIMENTAL DETAILS

### A. Samples and Measurement Methods

The irradiation resistance of nanofilm materials was investigated by the in-situ measurement of sensitivity of them-based Hall sensors during irradiation with neurons. For this purpose, the sensors' active elements of the Greek-cross shape with four leads were fabricated from nanofilms using photolithography, Fig. 1. Electrical contacts were created by the deposition of gold pads with thickness $(1.5 – 2.0)$ µm and subsequent bonding them (thermocompression method) with

This work has been carried out within the framework of (i) the EUROfusion Consortium and has received funding from the Euratom research and training programme 2014-2018 and 2019-2020 under grant agreement No 633053; (ii) the Graphene Flagship Consortium under grant agreement No 785219. The views and opinions expressed herein do not necessarily reflect those of the European Commission.

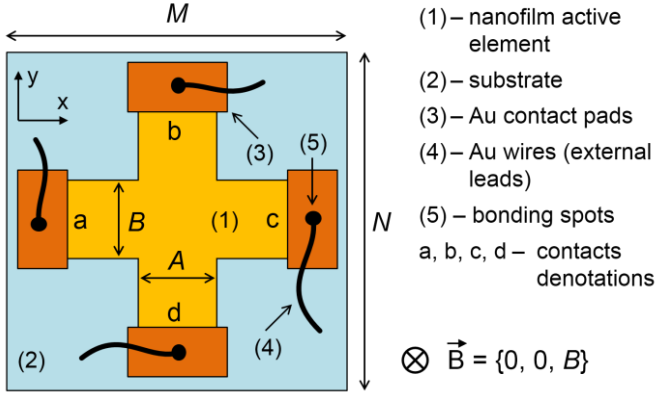

Fig. 1. The geometry of investigated Hall sensors (not to scale).

the external leads that were gold wires with a diameter of 30 µm.

In the magnetic field $\mathbf{B} = \{0, 0, B\}$, perpendicular to the sensor surface, Fig. 1, at transmitting through one pair of its opposite leads the DC current $I$, on the other pair the voltage $V$ is generated [6]:

$$V^{bd}(I^{ac}, B) = V_H^{bd}(I^{ac}, B) + V_0^{bd}(I^{ac}), \quad (1)$$

$$V_H^{bd}(I^{ac}, B) = (R_H/d) \cdot I^{ac} \cdot B \quad (2)$$

where the upper indices in $I$ and $V$ denote the leads, Fig. 1, to which the positive (first index) and negative (second index) terminals of the current source and voltmeter respectively are connected to; $V_H$ is the Hall voltage that depends on $B$; $V_0$ is the parasitic residual (offset) voltage, that arises due to the non-ideality of the active element shape and/or asymmetry in the bonding of external leads, and is independent on $B$; $R_H$ is the Hall coefficient of the material of active element; $d$ is its thickness.

Hall voltage $V_H$ determines the main functional characteristic of sensor – the current-related sensitivity to magnetic field:

$$S = |V_H^{bd}(I^{ac}, B) / (I^{ac} \cdot B)| = R_H/d, \quad (3)$$

which increases with decreasing of the active element thickness $d$, as can be seen from (2). Therefore, it is advisable to use for Hall sensors the films with nanoscale thickness $d$.

As can be seen from (2) and (3), the stability of the sensor under radiation conditions is determined, first of all, by the Hall coefficient $R_H$, which is inversely proportional to the concentration of free charge carriers $n$: $R_H \sim 1/n$ [6].

Minimizing the offset $V_0$ in (1) is the main task to improve the Hall sensors accuracy. In order to solve this problem, within this work the spinning-current method was used, which is based on the fact that, for an ideally symmetric active element, at synchronous re-switching of the voltmeter and the current source (so called "spinning" of current) $V_H$ and $V_0$ change the sign in different ways but preserve the absolute value [6]:

$$V_H^{bd}(I^{ac}, B) = V_H^{ca}(I^{bd}, B), V_0^{bd}(I^{ac}) = -V_0^{ca}(I^{bd})$$

Thus, an averaging of four values of the sensor output voltage $V$ (1), obtained at complete spinning of current relative to sample, allows to minimize the $V_0$ influence and to obtain Hall voltage:

$$V_H(I, B) = (1/4) \cdot [V^{bd}(I^{ac}, B) + V^{ca}(I^{bd}, B) +$$
$$+ V^{db}(I^{ca}, B) + V^{ac}(I^{db}, B)]. \quad (4)$$

As studies shown, the spinning-current method can reduce the offset by ~ $10^4$ times.

*B. Irradiation Testing*

Experiments on neutron irradiation were carried out in the research nuclear reactors LVR-15 (the Nuclear Physics Institute of the Czech Academy of Sciences) and IBR-2 (the Joint Nuclear Research Institute). The LVR-15 is a reactor with water-cooled core that allows to obtain neutron fluxes in which the number of thermal neutrons is an order of magnitude greater than the number of fast ones. The irradiation in this reactor was used for the nuclear doping of indium-containing nanofilms in order to increase their irradiation resistance.

The IBR-2 core is cooled by liquid sodium, providing a flux with ~ 40% of fast neutrons. In addition, due to the use of the reactivity's mechanical modulation, a neutron flux intensity in the irradiation channels of this reactor exceeds $10^{17}$ n·m$^{-2}$·s$^{-1}$ [7]. These facts approximate the testing conditions in IBR-2 to the radiation fields where magnetic field sensors reside in fusion reactors. Therefore, main experiments on an investigation of the nanofilm sensors' irradiation resistance were carried out in IBR-2.

Neutron fluxes lead to the activation of materials, which significantly limits access to samples after irradiation, especially at high neutron fluences. Therefore, the nanofilm sensors' sensitivity was measured in-situ during irradiation with the using of special control electronics. The general scheme of the experiment is shown in Fig. 2a.

In order to allocate sensors in neutron fluxes, it was manufactured the remote heads based on the irradiation- and thermal-stable ceramics MACOR, Fig. 2b. Each head contained up to 4 Hall sensors; a miniature solenoid, providing the test magnetic field $B \approx 7$ mT, as well as a temperature sensor. A duralumin shield was used for the mechanical protection of the head.

The control electronics managed the operation of solenoid and temperature sensor, provided a power supply and a measurement of output voltage for Hall sensors, an implementation of the spinning-current algorithm and the $V_H$ determination according to (4). The obtained data were transmitted to a local computer with special software, where sensitivity $S$ was calculated according to (3), with the further storing of results on a remote server. For electrical communications, the irradiation-resistant cables type of "twisted pair" were used in the system.

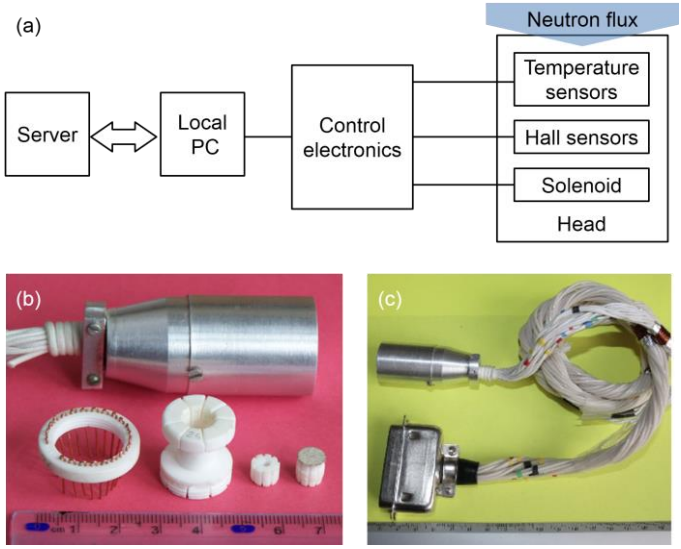

Fig. 2. The general scheme of experiment (a), the remote head's components (b), and the head with samples in assembly (c).

## III. RESULTS AND DISCUSSION

### A. Semiconductor Nanofilms

Semiconductors are the most common materials for Hall sensor due to their high $R_H$ value and well-developed production technology. However, as already noted, devices based on silicon are not capable to operate under high neutron fluxes for a long time. This limitation can be overcome to some extent by using indium-containing III-V semiconductors such as indium arsenide and indium antimonide. Increased irradiation resistance of such materials has been proven in previous studies of authors, that were carried out on the single-crystal whiskers [8].

Neutron irradiation changes the charge carrier concentration $n$ in InAs and InSb for account of two processes – the formation of radiation defects (predominantly by fast neutrons) and the transmutation doping (predominantly by thermal neutrons) according to the reaction:

$$\text{In115 (n, γ) In116m (β}^-\text{) Sn116,} \quad (5)$$

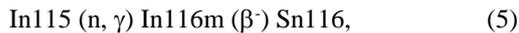

The carrier concentration's change rate in the dependence on fluence can be represented as:

$$\Delta n/\Delta F \approx \alpha - \beta \cdot n_0, \quad (6)$$

where $n_0$ is the initial charge carrier concentration, $\Delta n$ is its change under action of irradiation, $\Delta F$ is the neutron fluence increment, $\alpha$ is the coefficient of impurity injection due to the nuclear doping, $\beta$ is the cross section for the radiation defects formation [8]. It is clear that $\alpha$ and $\beta$ depend on the neutron energy spectrum in a particular reactor.

As can be seen from (6), the carrier concentration (and, accordingly, the Hall coefficient) in the sensor material during irradiation will be stable at $\alpha \approx \beta \cdot n_0$. This condition can be achieved by chemical (during semiconductor nanofilm growth) and nuclear (via the irradiation in reactor with a high content of thermal neutrons) doping to obtain the required concentration value $n_0$.

Semiconductor sensors with the sizes of active element $A = B = 200$ µm and of chip $M = N = 1$ mm were made on the basis of single-crystal nanofilms InAs and InSb with the thickness $d = 100$ nm, which were grown epitaxially on the semi-insulating gallium arsenide substrates of 400-µm thickness with the formation of InAs/i-GaAs and InSb/i-GaAs heterostructures. InAs nanofilms were grown using Metal-Organic Chemical Vapor Deposition (MOCVD) method, InSb – using the Vacuum Epitaxy method. During growth, the material was chemically doped with atoms of silicon (for InAs) and tin (for InSb). Nuclear doping of the samples was carried out in the radiation fluxes of the LVR-15 reactor.

Fig. 3 shows dependencies of sensitivity on fluence $S(F)$ for two sensors based on InAs/i-GaAs with initial carrier concentration $n_0 = 2.4 \cdot 10^{18}$ cm$^{-3}$ and $n_0 = 6.0 \cdot 10^{18}$ cm$^{-3}$. As can be seen, the initial concentration's optimization can significantly improve the sensors stability without significant sensitivity loss. In so doing, it was possible to obtain samples that retain the $S$ value up to $10^{22}$ n·m$^{-2}$, which corresponds to the level of radiation loads on Hall sensors in the ITER reactor.

### B. Metal Nanofilms

Due to the relatively small $R_H$ value and the low sensitivity $S$, metals are not common materials for use in Hall sensors. However, their advantages, such as high irradiation and thermal resistance, make them promising for use in sensors which are applied in most severe radiation environment.

In this work, it was investigated sensors based on the nanofilms of gold, which is chemically stable, has a low $1/f$-

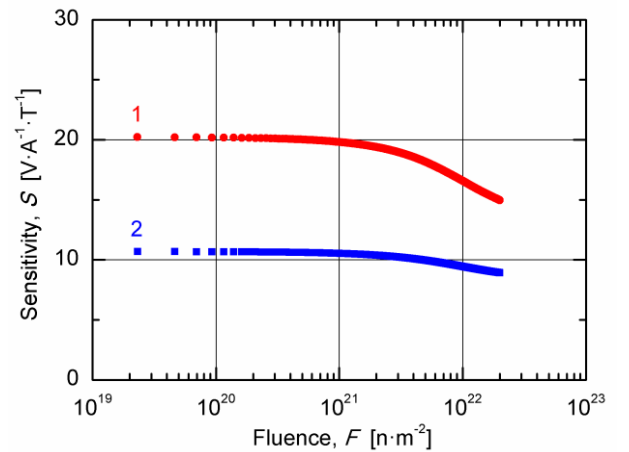

Fig. 3. Dependence of the sensitivity on the neutron fluence for Hall sensors based on InAs/i-GaAs nanogeterostructures with carrier concentration $n_0 = 2.4 \cdot 10^{18}$ cm$^{-3}$; (curve 1) and $n_0 = 6.0 \cdot 10^{18}$ cm$^{-3}$ (curve 2).

noise, a weak temperature dependence of the $R_H$ coefficient, and is convenient for deposition of the high-quality thin films.

Gold sensors with the sizes of active element $A = B = 200$ µm and of chip $M = N = 2$ were made of nanofilms with thickness $d = (50-70)$ nm, which were deposited by two methods: Electron Beam Evaporation (EBE) and Vacuum Thermal Deposition (VTD). As the substrate, it was used sapphire ($Al_2O_3$), which has high thermal conductivity and irradiation resistance. Preliminary, a titanium adhesive layer with a thickness of (5-7) nm was deposited onto substrate in order to improve adhesion, as well as a platinum barrier layer of 5-nm thick was deposited in order to weaken the diffusion processes between Ti and Au, what is of particular importance at high temperatures (> 200 °C) [9]. Studies of other authors have shown that the Ti/Au/Pt metallization system is stable when heated up to 500 °C [10].

In-situ sensitivity dependences on fluence at the irradiation up to DEMO-relevant fluence $F \approx 10^{24}$ n·m$^{-2}$ are shown in Fig. 4 for two samples manufactured by different methods. As can be seen, the samples sensitivity $S$ remains almost unchanged. Peak-like spikes are not connected with neutrons impact because the $S$ values are the same before and after they occur. More probable origin is some technological processes in the reactor (the launching of powerful equipment, the maneuvering by power, etc.), which affected the measuring system through electromagnetic interferences and/or magnetic field generation.

In general, for all investigated samples, the difference in $S$ values at the beginning and finishing of irradiation does not exceed 3%, which is within the measurement error. Moreover, the obtained dependencies have no any peculiarities that would indicate the degradation of the gold sensors upon further irradiation. This confirms the prospect of the gold nanofilms application under the severe environment of the new-generation fusion reactors type of ITER and DEMO.

*C. Single-Layer Graphene*

Single-layered graphene has the lowest possible thickness (only one layer of atoms, $d \sim 0.34$ nm) and low carrier concentration $n$, which allows to create Hall sensors with exceptionally high sensitivity to magnetic field [11]. Another advantage of graphene is the high resistance to corpuscular radiation, which has been confirmed in many studies related to the charged-particle irradiation [12-13]. However, experimental data on the neutron fluxes effect on single-layer graphene are almost absent in the literature. In this work, the authors conducted the world's first testing of the graphene Hall sensors in neutron fluxes.

High irradiation resistance of graphene is explained by low probability of the bombarding particles' collision with its atoms, which are arranged in a two-dimensional hexagonal structure. As the calculations show, for neutrons this probability is several orders of magnitude lower than for ions and consists only $\sim 10^{-5}$ [14]. In addition, the absence of a bulk crystal structure makes it impossible to developing in graphene of the large-scale atomic displacements cascades and irradiation damages are local only. Another peculiarity is the graphene's ability to "self-heal" defects, which is realized through the re-ordering of the damaged structure and/or the adatoms capturing by vacancies, and is not observed in bulk materials [12]. The combination of these factors makes graphene promising for use in severe radiation environment.

Graphene sensors with the sizes of active element $A = B = 100$ µm and of chip $M = N = 1$ mm were manufactured in collaboration with the AMO GmbH Company (Germany) on the basis of the commercial graphene (Graphenea, Spain) that was grown on copper foil by the Chemical Vapor Deposition (CVD) method, and then was transferred to sapphire. In the process of the electrical contacts creation, a 20-nm nickel contact layer was spread additionally on graphene before the gold pads' deposition. In order to protect the graphene active elements against the atmosphere influence, a layer of $Al_2O_3$ encapsulation of 80-nm thick was deposited by the Atomic Layer Deposition (ALD) method.

Irradiation testing of graphene sensors in IBR-2 was carried out up to fluence $F \approx 1.5 \times 10^{20}$ n·m$^{-2}$. This restriction probably is connected with the imperfection of used technology of the metal-graphene electrical contacts, and, as a result, the contacts are broken due to the radiation heating. Further improvement of the technology of metal-graphene contacts will allow to enlarge the neutron fluence range.

Fig. 5 shows the in-situ dependence of sensitivity on fluence $S(F)$ for one of the samples. The difference between the sensitivity values at the beginning and at the end of irradiation is no more than 3 %, which indicates the sensors' high resistance to the neutron action within the considered fluence range.

On the other hand, the dependence $S(F)$ shows a slight nonlinearity in the central part, Fig. 5. To determine the causes of its appearance, the neutron irradiation effect on the graphene crystalline structure was investigated. For this purpose, it was used the samples without encapsulation layer and electrical contacts, for which the Raman spectra were compared before irradiation and after irradiation with fluences of $F \approx 4.6 \times 10^{19}$ n·m$^{-2}$ and $F \approx 4.1 \times 10^{20}$ n·m$^{-2}$. The last fluence value is ~ 3 times higher than the maximum fluence achieved at the in-situ measurements of sensitivity, Fig. 5. Raman spectra did not shown an appearance of the visible amount of structural defects in graphene after irradiation. Thus, the observed peculiarities on $S(F)$ are not related with the radiation damages to graphene. The reasons for their occurrence may be the effect of heating on the metal-graphene contacts, the appearance of mechanical stresses on the graphene from the side of the surrounding materials, etc.

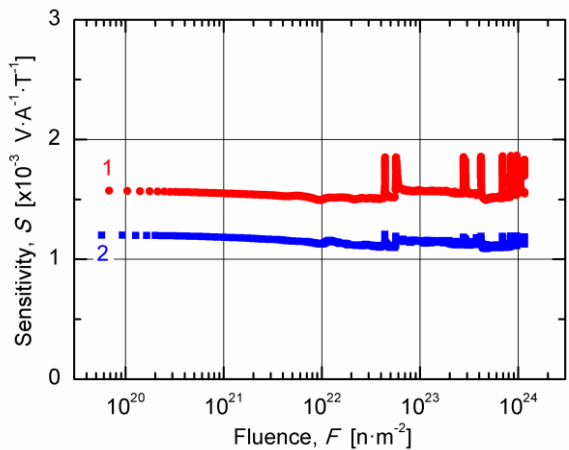

Fig. 4. Sensitivity dependence on neutron fluence for Hall sensors based on gold nanofilms, deposited by VTD (curve 1) amd EBE (curve 2) methods.

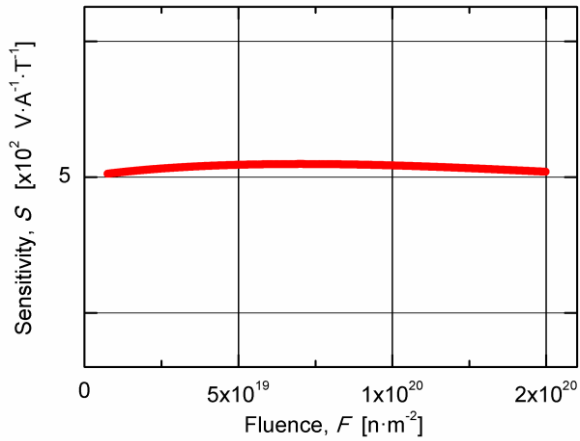

Fig. 5. Dependence of the sensitivity on the neutron fluence for Hall sensor based on single-layered graphene.

## CONCLUSION

In-situ measurements carried out during irradiation with reactor neutrons confirmed the prospect of using of different functional materials' nanofilms in devices of magnetic field measurement in radiation environment. Nanofilms of indium-containing III-V semiconductors provide high Hall sensitivity (tens $V \cdot A^{-1} \cdot T^{-1}$) and capable to operate up to fluences of $\sim 10^{22}$ $n \cdot m^{-2}$, which allows to use them in the plasma's magnetic diagnostics systems of the ITER-type fusion reactors. Gold nanofilms have a relatively low sensitivity (several $mV \cdot A^{-1} \cdot T^{-1}$), but it remains unchanged at least up to fluences of $\sim 10^{24}$ $n \cdot m^{-2}$ with prospects of stability preservation at even higher irradiation loads. This, together with the high thermal stability of the gold's electrophysical characteristics, makes gold nanofilms suitable for use in the plasma's diagnostics systems for the DEMO-type reactors. Graphene provides ultra-high sensitivity (hundreds of $V \cdot A^{-1} \cdot T^{-1}$) and has already shown the stability of its structure up to fluences of $\sim 4 \times 10^{20}$ $n \cdot m^{-2}$. However, for further in-situ measurements of parameters of the graphene sensors in neutron fluxes, it is necessary to resolve certain technological problems in order to improve their reliability and thermal stability.

## CONCLUDING REMARKS

This paper did not separately investigate an effect on the sensors sensitivity of impurities in their components (active elements, substrates, contact pads, communication lines), which may be created due to transmutation under ITER and DEMO environment. It is related to the fact that fast neutrons, which create radiation defects, are considered to be the main reason of the material parameters' change in fusion reactors. Transmutation, in turn, is mainly due to thermal neutrons, whose number is relatively small in thermonuclear reactors.

That is why the experiments on the sensors' irradiation testing were carried out in the fast-neutron reactor IBR-2, where the ratio of thermal and fast neutron fluxes is close to the case of fusion reactors. The observed sensitivity stability indicates that in the considered fluence ranges neither radiation defects nor transmutation impurities noticeably affect the sensors components. On the other hand, it is assumed that gold and graphene sensors are capable to tolerate significantly higher total neutron fluences than were achieved in this work. In so doing, the thermal neutron fluence can be high enough for noticeable transmutation. However, in the IBR-2 environment, it is impossible to achieve such irradiation loads in a short time. Therefore, the transmutation issue requires to carry out separate experiments in the thermal-neutron nuclear reactors using the methods of the activation and elemental analysis, which is beyond the scope of this paper, but is planned in the future.

ACKNOWLEDGMENT

The authors express their gratitude to I. Vasil'evskii for making gold sensors by the VTD method; H. Windgassen and S. Scholz for assistance in making of the electrical contacts and external leads bonding for graphene sensors; as well as M. Bulavin and S. Kulikov for carrying out the neutron irradiation in the IBR-2 reactor.